\documentclass[a4]{article}
\usepackage[left=2cm,right=2cm,top=2cm,bottom=2cm,bindingoffset=0cm]{geometry}
\usepackage{multirow}

\usepackage{amsfonts, amsmath, amssymb}
\usepackage{graphicx, float}
\usepackage{subcaption}

\newcommand{\be}{\begin{equation}}
\newcommand{\ee}{\end{equation}}
\newcommand{\Tr}{{\rm Tr}}

\graphicspath{{./pictures/}}

\renewcommand{\Re}{\text{Re}}
\renewcommand{\Im}{\text{Im}}

\begin{document}
	\title{ONE-LOOP CORRECTION CONTRIBUTIONS IN THE DECOUPLING LIMIT OF GENERALIZED YUKAWA MODEL}
	\author{M.S.~Dmytriiev$^1$, V.V.~Skalozub$^2$, \\ $^1$dmytrijev\_m@ffeks.dnu.edu.ua, $^2$Skalozubv@daad-alumni.de, \\
		Oles Honchar Dnipro National University, \\
		72, Gagarin Ave., Dnipro 49010, Ukraine}
	
	\maketitle
	
	\begin{abstract}
		We consider the generalized Yukawa model consisting of two Dirac fermions and two scalar fields. One of the scalars  is assumed to be much heavier than the other particles, so it decouples at low energies. Low-energy effective Lagrangian (EL) of the model is derived. It describes the contributions of the heavy scalar into observables in the  limit when this particle is decoupled. We consider the cross-sections of $s$- and $t$-channel processes, obtained within the complete model and its low-energy approximation. The contributions  of the one-loop corrections in the cross-sections coming from light particles  are analyzed. These are corrections to the parameters of the heavy boson and the contributions of the one-loop mixing of light and heavy scalars. We identify the ranges of  Yukawa's couplings where the corrections are significant. We find that if the interaction between fermions and either light or heavy scalar is strong enough,  the derived EL could not be applied for the description of the analyzed  cross-sections even if the heavy scalar decouples. Implications of our results in searching  for new particles beyond the Standard model are discussed.
		
	\end{abstract}
	\vspace{0.21cm}
	{\fontsize{9}{9}\selectfont Received 18.10.2023; Received in revised form 20.11.2023; Accepted 11.12.2023}
	\vspace{0.42cm}
	\section{Introduction}
	\label{sec:introduction}
	
	There exist many models of physics, which introduce new heavy particles beyond the Standard model (SM). Signals of such new states could be parametrized either within a complete model of new physics or its low-energy approximation. The latter consists of the effective Lagrangian (EL), which describes effects of interactions with heavy particles on the dynamics of light fields when heavy states decouple. Parameters of this EL should be constrained in experimental data analysis. These constraints, in turn, are translated into limits on parameters of a corresponding complete model. This approach is employed in many researches, for example, in \cite{bibl:cms-wilson-coefficients-limits, bibl:ellis}. In our investigation we show that the EL considered in this treatment is valid only if loop corrections coming from the light sector of a corresponding complete model are negligible in the decoupling limit.
	
	Our research is fulfilled  for two scattering processes which take place in $s$- and $t$-channel, respectively. Transformational properties of fields and complex interactions within the SM are inessential for our analysis. Hence, we consider generalized Yukawa model instead of a complete SM extended with some new states, similarly to \cite{bibl:brandt-loop-corrections-from-effective-vertexes, bibl:dmytriiev-skalozub-direct-search-of-dark-matter}. Our model consists of scalars $\phi$ and $\chi$ and fermions $\psi_1$ and $\psi_2$. We put $\chi$ boson to be much heavier than the other particles of the model, so it decouples at low energies. We consider scattering processes $\psi_1\bar{\psi}_1\rightarrow \psi_2\bar{\psi}_2$ and $\psi_1\psi_2\rightarrow \psi_1\psi_2$. We derive low-energy EL for this model. This EL approximates the contributions of the $\chi$ boson in observables when $\chi$ decouples. We work out the cross-sections of the considered reactions within the complete and the effective models. Contributions of radiative corrections in the  cross-sections at low energies are analyzed. These are corrections to the $\chi$ boson parameters and the contributions of the one-loop mixing of light and heavy scalars. We identify the ranges of  Yukawa's couplings where the radiative corrections are significant. We figure out that if it is so,  the EL could not be applied for  describing the reaction cross-sections at low energies. In particular, the loop corrections might be significant in the scenarios of strong couplings in new physics models discussed in the literature, for example, in \cite{bibl:ellis, bibl:criado, bibl:contino}. We argue that the contributions of the loop corrections in observables should be estimated when the constraints on EFT operator coefficients are derived from experimental data.
	
	We consider only the dimension-$6$ effective operators in the EL of our model. As it is known, the effective operators of dimension $5$ introduce a lepton-number violation \cite{bibl:cms-wilson-coefficients-limits}, which is strongly constrained by experiments. Low-energy EL  investigated in our paper does not contain such operators. So our model is aligned with the current experimental data.
	
	According to \cite{bibl:contino}, there are some scenarios when the contributions of effective operators of dimensions $> 6$ in observables are non-negligible or even leading. In our treatment these operators are significant at high energies just below the EFT UV cutoff, but at low energies they are suppressed by  heavy particle mass.  Hence, in our analysis we neglect effective operators of dimensions $> 6$.
	
	There exist investigations of applicability ranges of low-energy effective field theories \cite{bibl:contino, bibl:hartmann}. However, up to our knowledge, no detailed studies of the loop corrections  coming from light particle loops in the decoupling regime have been carried out so far.
	
	For example, it is suggested in \cite{bibl:hartmann} that the one-loop corrections should be taken into account in scattering amplitudes within a low-energy EFT to match precision of modern experimental measurements. Authors of this investigation consider the scattering amplitudes in a next-to-leading order in effective vertexes. In contrast below  we derive a number of limits on the parameters of the complete model, and do not consider the loops introduced by the effective interactions.
	
	Detailed analysis of EFT applicability range is performed in \cite{bibl:contino}. Particularly, it is stated therein  that if the interaction couplings are big, the loop corrections are to be significant at low energies. However, it is not specified  which values of couplings should be treated as big ones. In our research, we provide the estimates of such values for considered model and also identify the loop diagrams which  are the most significant  in various scenarios.
	
	This paper is organized as follows. In section\,\ref{sec:derivation-of-effective-lagrangian} we introduce our model and derive the low-energy EL for it. In sections\,\ref{sec:s-channel-process-analyzis} and \ref{sec:t-channel-process-analyzis} we analyze the contributions  of loop corrections in the $s$- and $t$-channels. Finally, we summarize  our results and conclusions in section\,\ref{sec:conclusion}.
	\section{The model}
	\label{sec:derivation-of-effective-lagrangian}
	
	Let us consider the generalized Yukawa model which consists of two Dirac fermions $\psi_1$ and $\psi_2$ and two scalar bosons $\phi$ and $\chi$. Fermions and scalars of the model interact via Yukawa's couplings. We assume that $\chi$ is much heavier than the other particles. The Lagrangian of the model reads:
	\begin{align}
	\label{full-model-lagrangian}
	\mathcal{L} &= \frac{1}{2}\left(\partial_{\mu}\phi\right)^2 - \frac{1}{2}\mu^2\phi^2 + \frac{1}{2}\left(\partial_{\mu}\chi\right)^2 - \frac{1}{2}\Lambda^2\chi^2 - \lambda\phi^4 - \rho\phi^2\chi^2 - \xi\chi^4 + \nonumber \\ 
	&+ \sum\limits_{a=1;2}\bar{\psi}_a\left(i\hat{\partial} - m_a - g_{\phi}\phi - g_{\chi}\chi\right)\psi_a.
	\end{align}
	Here $\mu$ and $\Lambda$ are masses of scalars $\phi$ and $\chi$, while $m_1$ and $m_2$ are masses of fermions $\psi_1$ and $\psi_2$, respectively. $\lambda$, $\rho$ and $\xi$ denote scalar self-interaction constants. $g_{\phi}$ and $g_{\chi}$ are the Yukawa couplings. All the parameters  are real.
	
	The scalar $\chi$ decouples when energies of scattering particles are much less than $\Lambda$. In the model proposed here $\Lambda\gg\mu;\,m_1;\,m_2$, so interactions of $\phi$, $\psi_1$ and $\psi_2$ could be described with a low-energy effective Lagrangian when $\chi$ boson decouples. To derive this EL, we assume that $\chi$ particles are absent in the initial and final states, and integrate out the heavy scalar:
	\begin{equation}
	\label{effective-lagrangian-definition}
	\exp\left(i\int d^4x \mathcal{L}_{eff}\right) = \int\mathcal{D}\chi \exp\left(i\int d^4x \mathcal{L}\right) = \int\mathcal{D}\chi e^{iS}.
	\end{equation}
	We calculate this functional integral in the Gaussian approximation, similarly to \cite{bibl:dmytriiev-skalozub-eff-lagrangian}. In this approximation action $S$ is expanded into  Taylor series around some given $\chi_0(x)$:
	\begin{align}
	\label{taylor-series-for-action}
	S &= S[\chi_0] + \int d^4x \frac{\delta S}{\delta\chi(x)}\Bigr|_{\chi=\chi_0}(\chi - \chi_0)(x) + \nonumber \\ 
	&+ \frac{1}{2}\int d^4x_1 d^4x_2 \frac{\delta^2 S}{\delta\chi(x_1)\delta\chi(x_2)}(\chi - \chi_0)(x_1)(\chi - \chi_0)(x_2) + O\left[(\chi - \chi_0)^3\right].
	\end{align}
	Then we put $\chi_0(x)$ such that $\frac{\delta S}{\delta \chi(x)}\Bigr|_{\chi = \chi_0} = 0$. Thus, $\chi_0(x)$ satisfies classical motion equation for $\chi$. Terms $O\left[(\chi - \chi_0)^3\right]$ in \eqref{taylor-series-for-action} are all proportional to scalar self-interaction couplings. We assume that these constants are so small that terms $O\left[(\chi - \chi_0)^3\right]$ could be neglected. Finally, $S$ is approximated with the following expression:
	\begin{equation}
	\label{gaussian-approximation-for-action}
	S \approx S[\chi_0] + \frac{1}{2}\int d^4x_1 d^4x_2 \frac{\delta^2 S}{\delta\chi(x_1)\delta\chi(x_2)}(\chi - \chi_0)(x_1)(\chi - \chi_0)(x_2).
	\end{equation}
	Equation for $\chi_0(x)$ reads:
	\begin{equation}
	\label{classical-motion-equation-for-chi}
	(\partial^2 + \Lambda^2)\chi_0 = -2\rho\phi^2\chi_0 - 4\xi\chi_0^3 - J_{\chi},\quad J_{\chi} = g_{\chi}\sum\limits_{a=1;2}\bar{\psi}_a\psi_a.
	\end{equation}
	We solve this equation employing the same approximations as in \cite{bibl:dmytriiev-skalozub-eff-lagrangian, bibl:belusca}. That is, we write down $\chi_0(x)$ as a series in growing powers of couplings:
	\[\chi_0 = \chi_0^{(0)} + \chi_0^{(1)} + \chi_0^{(2)} + ... \Rightarrow \begin{cases}
	(\partial^2 + \Lambda^2)\chi_0^{(0)} = -J_{\chi} \\
	(\partial^2 + \Lambda^2)\chi_0^{(1)} = -2\rho\phi^2\chi_0^{(0)} - 4\xi\left(\chi_0^{(0)}\right)^3 \\
	...
	\end{cases}.\]
	In the decoupling region we neglect the derivative term in \eqref{classical-motion-equation-for-chi}, since $|\partial^2\chi| \ll \Lambda^2|\chi|$. Hence, $\chi_0(x)$ is approximately equal to the following expression:
	\begin{equation}
	\label{classical-solution-for-chi}
	\chi_0 \approx \chi_0^{(0)} = -\frac{1}{\Lambda^2}J_{\chi}.
	\end{equation}
	Here we have omitted the other terms in the expansion for $\chi_0(x)$, since they are suppressed by higher powers of $\Lambda^{-2}$. We derive EL for our model only up to terms $O(\Lambda^{-4})$.
	
	We put \eqref{classical-solution-for-chi} to the expression for action $S$ and get the first term in \eqref{gaussian-approximation-for-action}:
	\begin{equation}
	\label{classical-action-for-chi}
	S[\chi_0] = \int d^4x \left[\frac{1}{2}\left(\partial_{\mu}\phi\right)^2 - \frac{1}{2}\mu^2\phi^2 - \lambda\phi^4 + \sum\limits_{a=1;2}\bar{\psi}_a(i\hat{\partial} - m_a - g_{\phi}\phi)\psi_a + \frac{1}{2\Lambda^2}J_{\chi}^2\right].
	\end{equation}
	For the second term in \eqref{gaussian-approximation-for-action} we have:
	\[\frac{\delta^2 S}{\delta\chi(x_1)\delta\chi(x_2)}\Bigr|_{\chi = \chi_0} = -\left[\partial^2 + \Lambda^2 + 2\rho\phi^2(x_1)\right]\delta(x_1 - x_2).\]
	We put this expression to \eqref{effective-lagrangian-definition} and integrate over the fluctuations $(\chi - \chi_0)(x)$. Eventually, we omit infinite constants and get the following expression for $\mathcal{L}_{eff}$:
	\[i\int d^4x \mathcal{L}_{eff} = iS[\chi_0] - \frac{1}{2}\Tr\ln{\left(\partial^2 + \Lambda^2 + 2\rho\phi^2\right)}.\]
	The second term here could be expanded in powers of $\rho$:
	\[\Tr\ln{\left(\partial^2 + \Lambda^2 + 2\rho\phi^2\right)} = \Tr\ln{(\partial^2 + \Lambda^2)} + 2\rho\int d^4x (\partial^2 + \Lambda^2)^{-1}(x;x)\phi^2(x) + O(\rho^2).\]
	The first term here is constant so we omit it. The second term describes radiative corrections to the $\phi$ mass from the loop with $\chi$ boson. This correction is absorbed in the renormalized value of $\phi$ mass. The other terms could be neglected, since they are suppressed either by higher powers of $\rho$ or $\Lambda^{-2}$.
	
	Finally, we get the following expression for $\mathcal{L}_{eff}$:
	\begin{equation}
	\label{effective-lagrangian}
	\mathcal{L}_{eff} = \frac{1}{2}\left(\partial_{\mu}\phi\right)^2 - \frac{1}{2}\mu^2\phi^2 - \lambda\phi^4 + \sum\limits_{a=1;2}\bar{\psi}_a(i\hat{\partial} - m_a - g_{\phi}\phi)\psi_a + \frac{g_{\chi}^2}{2\Lambda^2}\left(\bar{\psi}_1\psi_1 + \bar{\psi}_2\psi_2\right)^2.
	\end{equation}
	The last term here corresponds to non-renormalizable contact four-fermion interactions. Since this term is suppressed by $\Lambda^{-2}$, these interactions vanish if we put $\Lambda\rightarrow\infty$.
	
	The values of the parameters $m_1$, $m_2$ and $\mu$ are fixed in our analysis. These values are shown in table\,\ref{tab:fixed-parameters-values}.
	\begin{table}[h]
		\caption{Values of the model parameters which are fixed in the analyzis}
		\label{tab:fixed-parameters-values}
		\centering
		\begin{tabular}{c|c|c}
			$m_1$ & $m_2$ & $\mu$ \\
			\hline
			$0.511\,MeV$ & $105.658\,MeV$ & $5\,GeV$
		\end{tabular}
	\end{table}
	
	In the next two sections \ref{sec:s-channel-process-analyzis} and \ref{sec:t-channel-process-analyzis} we identify ranges of the model parameters where loop corrections are significant when $\chi$ decouples. In these scenarios \eqref{effective-lagrangian} could not be applied to estimate cross-sections of the analyzed processes at low energies. In our discussion we provide ranges of $\Lambda$ values for which we investigated contributions of radiative corrections.
	
	\section{The $s$-channel scattering process}
	\label{sec:s-channel-process-analyzis}
	
\begin{figure}[h]
		\centering
		\includegraphics[scale=0.4]{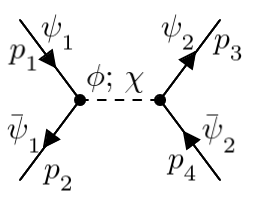}
		\caption{Diagram of the reaction $\psi_1\bar{\psi}_1\rightarrow \psi_2\bar{\psi}_2$ within the UV-complete model described by the Lagrangian \eqref{full-model-lagrangian}}
		\label{fig:s-channel-process-uv-complete-diagram}
	\end{figure}

\begin{figure}
		\centering
		\begin{subfigure}{0.5\textwidth}
			\centering
			\includegraphics[scale=0.35]{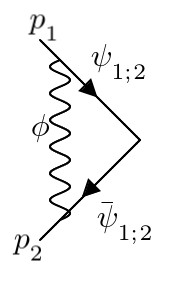}
			\caption{$\delta\Gamma_{\phi;\chi}^{(\phi)}(s;m_{12})$}
			\label{fig:phi-loop-in-yukawa-vertex}
		\end{subfigure}%
		\begin{subfigure}{0.5\textwidth}
			\centering
			\includegraphics[scale=0.35]{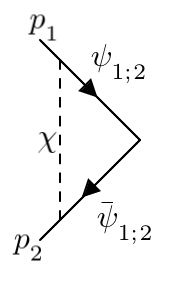}
			\caption{$\delta\Gamma_{\phi;\chi}^{(\chi)}(s;m_{12})$}
			\label{fig:chi-loop-in-yukawa-vertex}
		\end{subfigure}
		\caption{Diagrams of loop corrections to Yukawa vertexes in the model}
		\label{fig:yukawa-vertex-corrections-uv-complete-model}
	\end{figure}

	Let us consider the scattering process $\psi_1\bar{\psi}_1\rightarrow \psi_2\bar{\psi}_2$, which takes place in $s$-channel only. The diagram of this process is shown in Fig.\,\ref{fig:s-channel-process-uv-complete-diagram}. We derive matrix element of this process in the improved Born approximation. The latter implies that all one-loop radiative corrections are taken into account except for box diagrams. It was shown in \cite{bibl:dmytriiev-skalozub-direct-search-of-dark-matter} that contribution of box diagrams in the considered $s$-process cross-section within the model \eqref{full-model-lagrangian} is less than $1\,\%$ of the contribution of one-particle-reducible diagrams. Thus, we omit boxes in our treatment.
	Cross-section of this reaction within the UV-complete model \eqref{full-model-lagrangian} in the center-of-mass reference frame reads:
	\begin{align}
	\label{s-channel-process-uv-complete-cross-section}
	\sigma(\psi_1\bar{\psi}_1\rightarrow \psi_2\bar{\psi}_2) &= \Phi^{(s)}(s)\left|\begin{pmatrix}
	-ig_{\phi}\Gamma_{\phi}(s;m_2) \\
	-ig_{\chi}\Gamma_{\chi}(s;m_2)
	\end{pmatrix}^T\right.\times \nonumber \\ 
	&\times\left.\begin{pmatrix}
	s - \mu^2 - \Pi_{\phi\phi}(s) & -\Pi_{\phi\chi}(s) \\
	-\Pi_{\phi\chi}(s) & s - \Lambda^2 - \Pi_{\chi\chi}(s)
	\end{pmatrix}^{-1}\begin{pmatrix}
	-ig_{\phi}\Gamma_{\phi}(s;m_1) \\
	-ig_{\chi}\Gamma_{\chi}(s;m_1)
	\end{pmatrix}\right|^2, \\
	\Phi^{(s)}(s) &= \frac{s}{16\pi}\left(1 - \frac{4m_2^2}{s}\right)^{\frac{3}{2}}\sqrt{1 - \frac{4m_1^2}{s}},\quad s = (p_1 + p_2)^2, \nonumber \\ 
	\Gamma_{\phi;\chi}(s;m_{12}) &= 1 + \delta\Gamma_{\phi;\chi}^{(\phi)}(s;m_{12}) + \delta\Gamma_{\phi;\chi}^{(\chi)}(s;m_{12}).
	\end{align}
	Hereafter all cross-sections are averaged over spins of initial fermions and summed over spins of fermions in a final state. Quantities $\delta\Gamma_{\phi}^{(\phi)}(s;m_{12})$, $\delta\Gamma_{\phi}^{(\chi)}(s;m_{12})$ and $\delta\Gamma_{\chi}^{(\phi)}(s;m_{12})$, $\delta\Gamma_{\chi}^{(\chi)}(s;m_{12})$ in \eqref{s-channel-process-uv-complete-cross-section} denote radiative corrections to Yukawa interaction with $\phi$ and $\chi$, respectively. Second argument of these functions is a mass of the fermionic field involved in the corresponding vertex. Diagrams of these corrections are displayed in Fig.\,\ref{fig:yukawa-vertex-corrections-uv-complete-model}.
	We impose the following renormalization conditions on $\Gamma_{\phi}(s;m_{12})$ and $\Gamma_{\chi}(s;m_{12})$:
	\begin{equation}
	\label{renormalization-conditions-of-yukawa-vertexes}
	\Gamma_{\phi}(\mu^2;m_{12}) = 1,\quad \Gamma_{\chi}(\Lambda^2;m_{12}) = 1.
	\end{equation}
	Corrections $\delta\Gamma_{\phi;\chi}^{(\phi)}(s;m_{12})$ and $\delta\Gamma_{\phi;\chi}^{(\chi)}(s;m_{12})$ are calculated numerically with LoopTools software \cite{bibl:hahn}. Function $\Phi^{(s)}(s)$ in \eqref{s-channel-process-uv-complete-cross-section} is a kinematical factor. It is a product of contributions from integration over the momentum space of final particles and averaging and summation over spins orientations of initial and final particles.
	
\begin{figure}[h]
		\centering
		\begin{subfigure}{0.3\textwidth}
			\centering
			\includegraphics[scale=0.4]{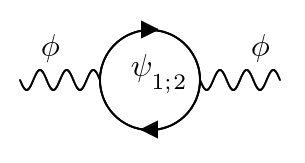}
			\caption{$\Pi_{\phi\phi}(s)$}
			\label{fig:phi-phi-polarization-operator}
		\end{subfigure}%
		\begin{subfigure}{0.3\textwidth}
			\centering
			\includegraphics[scale=0.4]{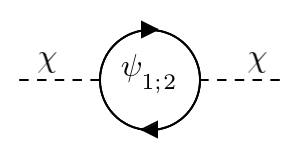}
			\caption{$\Pi_{\chi\chi}(s)$}
			\label{fig:chi-chi-polarization-operator}
		\end{subfigure}%
		\begin{subfigure}{0.3\textwidth}
			\centering
			\includegraphics[scale=0.4]{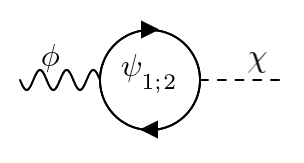}
			\caption{$\Pi_{\phi\chi}(s)$}
			\label{fig:phi-chi-polarization-operator}
		\end{subfigure}
		\caption{Radiative corrections in the two-point Green functions of scalar fields}
		\label{fig:radiative-corrections-to-scalar-propagators}
	\end{figure}

	$\Pi_{\phi\phi}(s)$, $\Pi_{\phi\chi}(s)$ and $\Pi_{\chi\chi}(s)$ in \eqref{s-channel-process-uv-complete-cross-section} denote radiative corrections in propagators of scalar fields. They are contributed by the diagrams shown in Fig.\,\ref{fig:radiative-corrections-to-scalar-propagators}.
	$\Pi_{\phi\phi}(s)$ and $\Pi_{\chi\chi}(s)$ describe loop corrections to the masses of $\phi$ and $\chi$ bosons, accordingly. $\Pi_{\phi\chi}(s)$ corresponds to one-loop mixing of $\phi$ and $\chi$. We calculate these corrections analytically and impose the following renormalization conditions on them:
	\begin{align}
	\label{renormalization-conditions-of-polarization-operator}
	\Re\,\Pi_{\phi\phi}(\mu^2) &= 0,\quad \Re\,\Pi_{\chi\chi}(\Lambda^2) = 0,\quad \Re\,\Pi_{\phi\chi}(\kappa^2) = 0, \nonumber \\
	\frac{\partial \Re\,\Pi_{\phi\phi}(s)}{\partial s}\Bigr|_{s = \mu^2} &= 0,\quad \frac{\partial \Re\,\Pi_{\chi\chi}(s)}{\partial s}\Bigr|_{s = \Lambda^2} = 0,\quad \frac{\partial \Re\,\Pi_{\phi\chi}(s)}{\partial s}\Bigr|_{s = \kappa^2} = 0.
	\end{align}
	Here $\kappa^2$ is an arbitrary renormalization scale. We put $\kappa = 1\,GeV$. One-loop contributions from scalar self-interactions in $\Pi_{\phi\phi}(s)$ and $\Pi_{\chi\chi}(s)$ are absorbed in renormalized values of $\phi$ and $\chi$ masses.
	
\begin{figure}[h]
		\centering
		\includegraphics[scale=0.5]{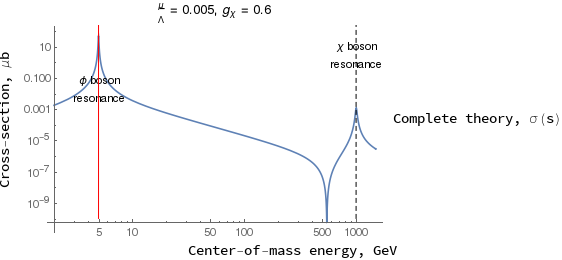}
		\caption{$\sigma$ as a function of the center-of-mass energy of $\psi_1$ and $\bar{\psi}_1$. Hereafter red solid line and black dashed line in the plots mark values of masses of $\phi$ and $\chi$ bosons, respectively}
		\label{fig:typical-s-process-cross-section-complete-theory}
	\end{figure}

	Example graph of $\sigma(s)$ dependency on the center-of-mass energy $\sqrt{s}$ is shown in Fig.\,\ref{fig:typical-s-process-cross-section-complete-theory}.
	As could be seen in Fig.\,\ref{fig:typical-s-process-cross-section-complete-theory}, $\sigma(s)$ develops two maximums, which correspond to the masses of $\phi$ and $\chi$. Dip between these maximums is introduced by interference of $\phi$ and $\chi$ exchange amplitudes in the expression of $\sigma(s)$ and one-loop mixing of these two fields. 
	
	It is worth noting that in this section we provide cross-sections plots in the logarithmic scale, so particles resonances shapes are distorted by the scale transformation.
	
	We derive approximate expression for \eqref{s-channel-process-uv-complete-cross-section} in the limit $\Lambda^2 \gg s$. If $\chi$ boson is very heavy, then we neglect its loop corrections to Yukawa vertexes, so the diagram in Fig.\,\ref{fig:chi-loop-in-yukawa-vertex} is omitted and $\lim\limits_{s \ll \Lambda^2}\delta\Gamma_{\phi;\chi}^{(\chi)}(s;m_{12}) \rightarrow 0$. $\sigma(s)$ in the limit of large $\Lambda$ reads:
	\begin{align}
	\label{s-channel-process-uv-complete-cross-section-large-lambda}
	\sigma(\psi_1\bar{\psi}_1\rightarrow\psi_2\bar{\psi}_2)\Bigr|_{\Lambda^2\gg s}&\approx \Phi^{(s)}(s)\left|\frac{(-ig_{\phi})^2\Gamma_{\phi}^{(eff)}(s;m_2)\Gamma_{\phi}^{(eff)}(s;m_1)}{s - \mu^2 - \Pi_{\phi\phi}(s)} +\right. \nonumber \\ 
	&+\left. \delta\mathcal{M}^{(mix)}(s) + \delta\mathcal{M}^{(4-ferm)}(s)\right|^2 = \sigma_{approx}(s), \\ 
	\Gamma_{\phi;\chi}^{(eff)}(s;m_{12}) &= 1 + \delta\Gamma_{\phi;\chi}^{(\phi)}(s;m_{12}), \nonumber \\ 
	\delta\mathcal{M}^{(4-ferm)}(s) &= \frac{g_{\chi}^2\Gamma_{\chi}^{(eff)}(s;m_1)\Gamma_{\chi}^{(eff)}(s;m_2)}{\Lambda^2 + \Pi_{\chi\chi}(s)}, \nonumber \\ 
	\delta\mathcal{M}^{(mix)}(s) &= \frac{2g_{\phi}g_{\chi}\Pi_{\phi\chi}(s)}{\left(\Lambda^2 + \Pi_{\chi\chi}(s)\right)\left(s - \mu^2 - \Pi_{\phi\phi}(s)\right)}. \nonumber
	\end{align}
	Here we omitted terms of higher orders in Yukawa couplings. There are three terms in the squared modulus factor in \eqref{s-channel-process-uv-complete-cross-section-large-lambda}. The first one corresponds to the s-channel reaction involving only light particles of the model. The second term $\delta\mathcal{M}^{(mix)}(s)$ is proportional to $\Pi_{\phi\chi}(s)$ and it describes contribution of one-loop mixing of scalar fields in $\sigma(s)$. Finally, the third term $\delta\mathcal{M}^{(4-ferm)}$ corresponds to the four-fermion interaction in the low-energy EFT. Unlike four-fermion vertex in \eqref{effective-lagrangian}, $\delta\mathcal{M}^{(4-ferm)}(s)$ contains polarization operator of $\chi$ boson $\Pi_{\chi\chi}(s)$. Since $\Pi_{\chi\chi}(s)\sim g_{\chi}^2$, $|\Pi_{\chi\chi}(s)|$ is significant if $|g_{\chi}|$ is big. Loop corrections to the Yukawa vertexes of $\chi$ boson also enter $\delta\mathcal{M}^{(4-ferm)}(s)$ in functions $\Gamma_{\chi}^{(eff)}(s;m_1)$ and $\Gamma_{\chi}^{(eff)}(s;m_2)$. These corrections emerge only from loops of light boson $\phi$ and their contributions are proportional to $g_{\phi}^2$. Hence, we should take them into account if interactions within light sector of the model are powerful enough.
	
\begin{figure}
		\begin{center}
			\begin{tabular}{cc}
				\includegraphics[scale=0.4]{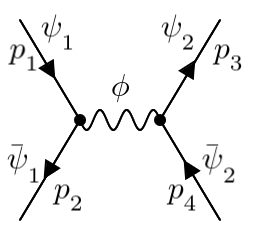} & \includegraphics[scale=0.4]{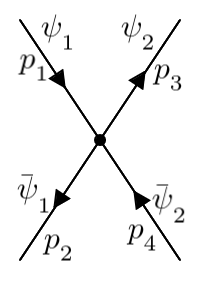}
			\end{tabular}
		\end{center}
		\caption{Diagrams which contribute cross-section of the $s$-channel reaction in the low-energy effective theory}
		\label{fig:s-channel-process-diagrams}
	\end{figure}

	Now we turn to the cross-section of the process $\psi_1\bar{\psi}_1\rightarrow \psi_2\bar{\psi}_2$ within the effective theory \eqref{effective-lagrangian}. Diagrams which are taken into account in this case are shown in Fig.\,\ref{fig:s-channel-process-diagrams}.
	Left diagram in Fig.\,\ref{fig:s-channel-process-diagrams} emerges from interactions of light fields only. We derive matrix element of this diagram in the improved Born approximation. Right diagram in Fig.\,\ref{fig:s-channel-process-diagrams} corresponds to the four-fermion contact interaction, which is specific to the low-energy EL \eqref{effective-lagrangian}. Cross-section of the reaction $\psi_1\bar{\psi}_1\rightarrow \psi_2\bar{\psi}_2$ within the low-energy effective theory is as follows:
	\begin{align}
	\label{s-channel-process-cross-section-effective-theory}
	\sigma_{eff}(\psi_1\bar{\psi}_1\rightarrow &\psi_2\bar{\psi}_2) = \Phi^{(s)}(s)\left|\frac{(-ig_{\phi})^2\Gamma_{\phi}^{(eff)}(s;m_1)\Gamma_{\phi}^{(eff)}(s;m_2)}{s - \mu^2 - \Pi_{\phi\phi}(s)} + \frac{g_{\chi}^2}{\Lambda^2}\right|^2.
	\end{align}
	Here we also take into account term $g_{\chi}^4\Lambda^{-4}$ despite the assumptions used in section \ref{sec:derivation-of-effective-lagrangian}. This term ensures that $\sigma_{eff}(s) \geq 0$. It could be observed that, contrary to \eqref{s-channel-process-uv-complete-cross-section-large-lambda}, radiative corrections to $g_{\chi}$ and $\Lambda$, as well as contribution of bosons one-loop mixing, are absent in \eqref{s-channel-process-cross-section-effective-theory}.
	
\begin{figure}
		\centering
		\includegraphics[scale=0.5]{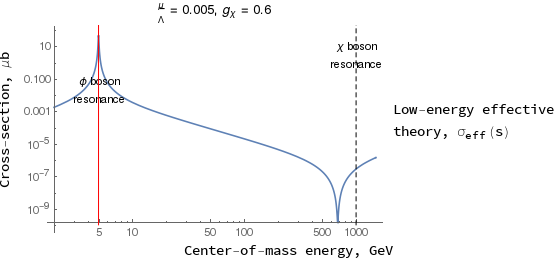}
		\caption{$\sigma_{eff}$ as a function of the center-of-mass energy of $\psi_1$ and $\bar{\psi}_1$}
		\label{fig:typical-s-process-cross-section-effective-theory}
	\end{figure}

	Example of $\sigma_{eff}(s)$ dependency on the center-of-mass energy $\sqrt{s}$ is shown in Fig.\,\ref{fig:typical-s-process-cross-section-effective-theory}. Similarly to $\sigma(s)$, $\sigma_{eff}(s)$ develops a maximum which corresponds to the mass of light $\phi$ boson. There is also a dip in $\sigma_{eff}(s)$ at $s > \mu^2$, which is introduced by interference between amplitudes of two diagrams in Fig.\,\ref{fig:s-channel-process-diagrams}.
	
\begin{figure}
		\centering
		\includegraphics[scale=0.4]{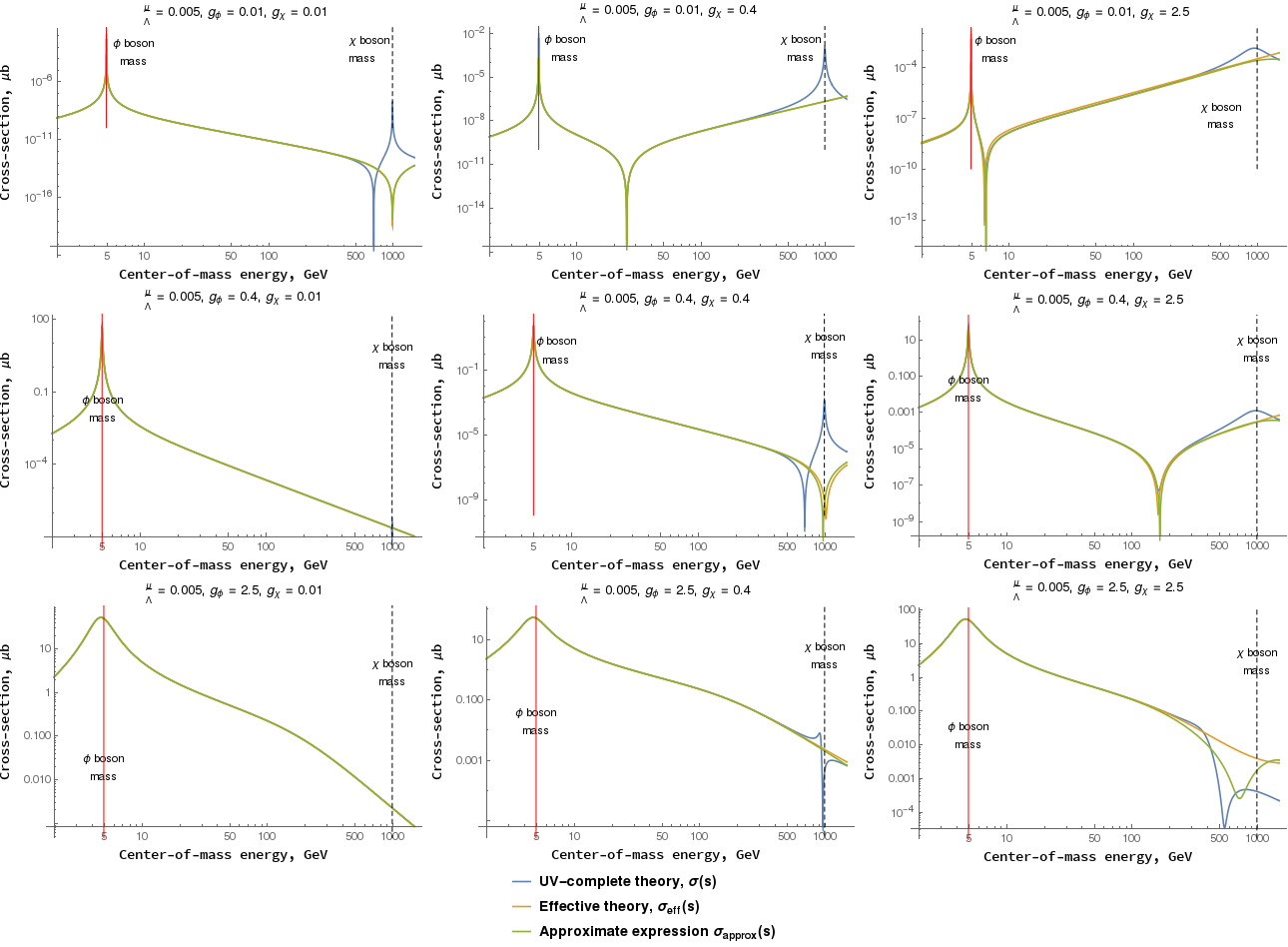}
		\caption{$\sigma(s)$, $\sigma_{eff}(s)$ and $\sigma_{approx}(s)$ at $\Lambda = 200\mu$ and different values of $g_{\phi}$ and $g_{\chi}$, $s$-channel process}
		\label{fig:s-process-cross-sections-within-complete-theory-and-low-energy-approximations}
	\end{figure}

\begin{figure}
		\centering
		\includegraphics[scale=0.4]{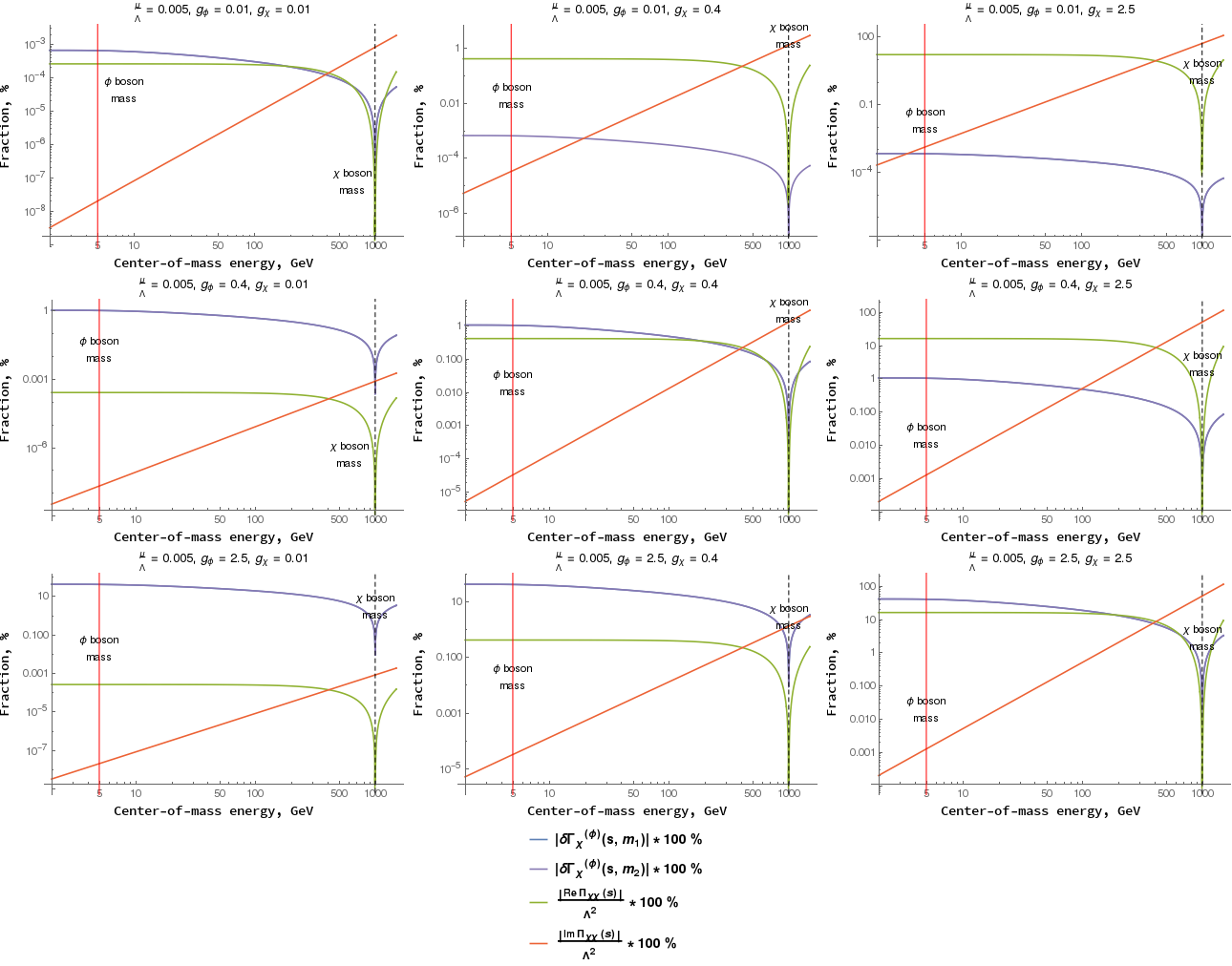}
		\caption{Values of loop corrections relative to corresponding parameters of the model at $\Lambda = 200\mu$ and different values of $g_{\phi}$ and $g_{\chi}$, $s$-channel process}
		\label{fig:s-process-loop-corrections-values}
	\end{figure}

\begin{figure}
		\centering
		\includegraphics[scale=0.4]{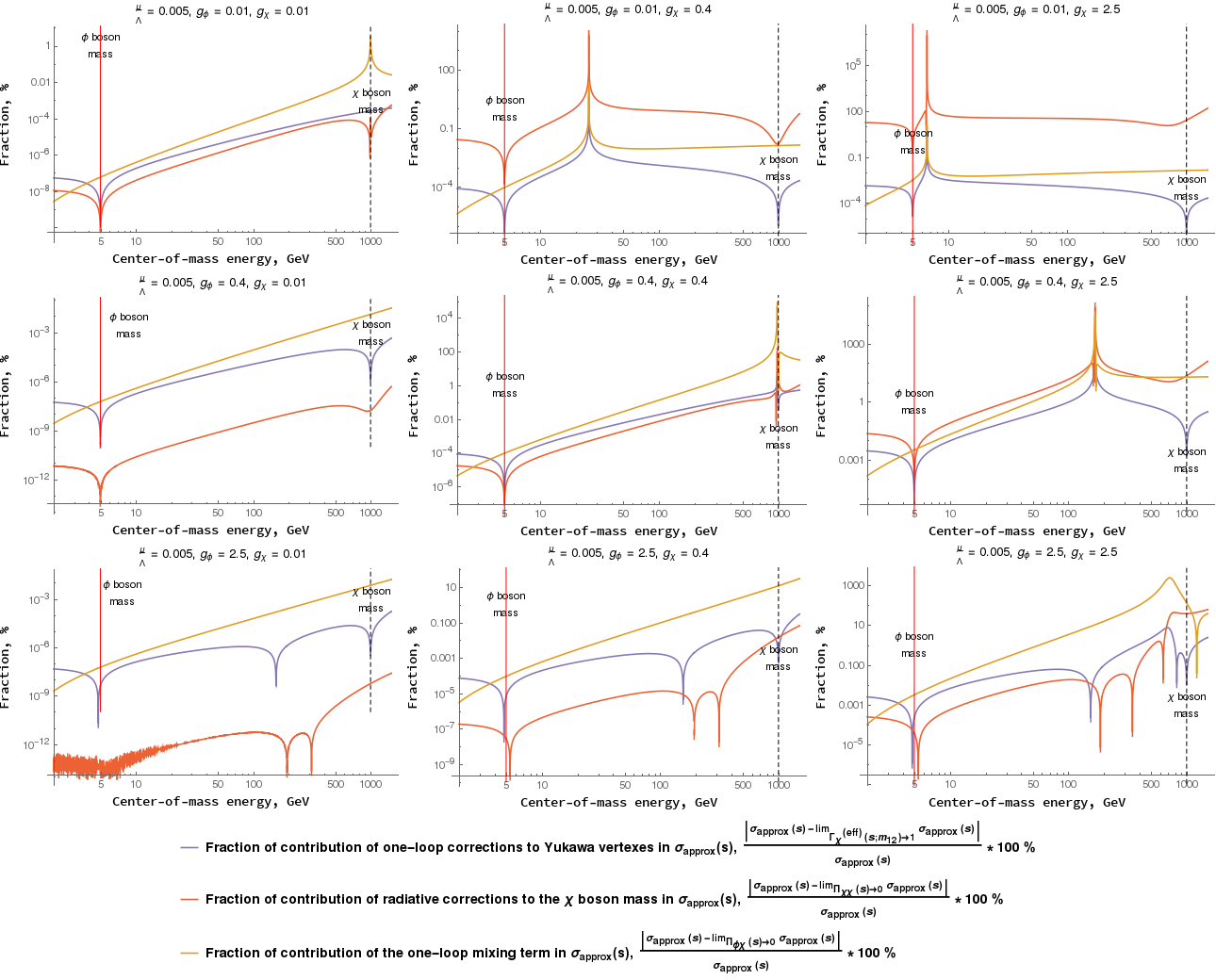}
		\caption{Contributions of loop corrections to Yukawa vertexes and mass of $\chi$ boson and one-loop mixing contribution in $\sigma_{approx}(s)$ at $\Lambda = 200\mu$ and different values of $g_{\phi}$ and $g_{\chi}$, $s$-channel process}
		\label{fig:s-process-loop-corrections-contributions}
	\end{figure}

	We plot $\sigma(s)$, $\sigma_{approx}(s)$ and $\sigma_{eff}(s)$ at different values of $\Lambda$, $g_{\phi}$ and $g_{\chi}$ as functions of $\sqrt{s}$ in Fig.\,\ref{fig:s-process-cross-sections-within-complete-theory-and-low-energy-approximations}. We also plot contribution of polarization operator of $\chi$ boson, contribution of radiative corrections to Yukawa vertexes of the latter and relative contribution of the one-loop mixing term in $\sigma_{approx}(s)$ in Fig.\,\ref{fig:s-process-loop-corrections-contributions}. Magnitudes of the first two corrections are also shown in the plots in Fig.\,\ref{fig:s-process-loop-corrections-values}. We analyze magnitudes and contributions of various radiative corrections at low energies and describe regions in the model parameters space where they are significant.
	
	We begin with the scenario when both $|g_{\phi}|$ and $|g_{\chi}|$ are small. If it is so, then $\Pi_{\chi\chi}(s)$ and one-loop mixing of scalar fields could be neglected in the reaction cross-section within the low-energy EFT, while $\Gamma_{\chi}^{(eff)}(s;m_{12})$ could be considered equal to $1$. For the model parameters values in table\,\ref{tab:fixed-parameters-values} we have that $|\delta\Gamma_{\chi}^{(\phi)}(s;m_{12})|\lesssim 0.01$, $|\Re\,\Pi_{\chi\chi}(s)|\lesssim 0.01\Lambda^2$ and $|\Im\,\Pi_{\chi\chi}(s)|\lesssim 0.01\Lambda^2$ if $|g_{\phi}| < 0.4$ and $|g_{\chi}| < 0.4$. These radiative corrections and one-loop mixing term contribute less than $1\,\%$ in $\sigma_{approx}(s)$ at $s \ll \Lambda^2$ for $20\mu \leq \Lambda \leq 200\mu$. Hence, if couplings of the model fermions to light and heavy scalars are small, then \eqref{effective-lagrangian} is applicable for the description of the reaction cross-section at low energies.
	
	Now we proceed to the scenarios when loop corrections in $\sigma(s)$ are significant and should be taken into account at low energies.
	
	It could be observed in the top right plot in Fig.\,\ref{fig:s-process-loop-corrections-contributions} that if $|g_{\chi}|$ is big, then contribution of radiative correction to $\Lambda$ in $\sigma_{approx}(s)$ is significant. For $|g_{\phi}|\approx 0.01$ and $2.5\leq |g_{\chi}| \leq 3$ $\Pi_{\chi\chi}(s)$ changes value of $\sigma_{approx}(s)$ on more than $10\,\%$ even at $s = O(\mu^2)$. In this limit all the other loop corrections are negligible, since they are proportional to $g_{\phi}^2$. $|\Pi_{\chi\chi}(s)|\gtrsim 0.1\Lambda^2$ for almost the whole range of energies $2m_2 < \sqrt{s} \leq \Lambda$ at $g_{\phi} \in [0.01;\,0.05]$ and $g_{\chi}\in [2.5;\,3]$ for a wide range of $\Lambda$ variation. Real and imaginary parts of $\Pi_{\chi\chi}(s)$ become dominant at low and high energies, respectively. We find out that for $0.01 \leq |g_{\phi}| \leq 0.05$ and $2.5\leq |g_{\chi}| \leq 3$ at low energies $\sqrt{s}\lesssim 0.5\Lambda$ $|\Re\,\Pi_{\chi\chi}(s)| > |\Im\,\Pi_{\chi\chi}(s)|$ and at higher energies $\sqrt{s}\gtrsim 0.5\Lambda$ $|\Im\,\Pi_{\chi\chi}(s)|$ dominates over $|\Re\,\Pi_{\chi\chi}(s)|$. If $|g_{\phi}| \ll |g_{\chi}|$ and $|g_{\chi}|$ is big, then contribution of heavy scalar in $\sigma(s)$ within the UV-complete theory \eqref{full-model-lagrangian} is significantly suppressed by $\Pi_{\chi\chi}(s)$. Thus, it is considerably overestimated in $\sigma_{eff}(s)$. According to the cross-sections plots in Fig.\,\ref{fig:s-process-cross-sections-within-complete-theory-and-low-energy-approximations}, $\sigma_{eff}(s)$ provides under- or overestimation of $\sigma(s)$ at different $s$. Hence, in the considered limit of couplings values and at $s = O(\mu^2)$ loop corrections to $\chi$ boson mass should be taken into account in the process cross-section at low energies.
	
	Contributions of loop corrections to Yukawa vertexes and one-loop mixing term in the process cross-section are small in the limit when $|g_{\phi}|\ll |g_{\chi}|$ and Yukawa couplings of $\chi$ are big. It follows from the top right plot in Fig.\,\ref{fig:s-process-loop-corrections-contributions} that if $g_{\phi} \in [0.01;\,0.05]$ and $g_{\chi}\in [2.5;\,3]$, then both contributions consist less than $1\,\%$ of $\sigma_{approx}(s)$ for most values of energies in the range $4m_2^2 < s \leq \Lambda^2$. According to the top right plot in Fig.\,\ref{fig:s-process-loop-corrections-values}, $|\delta\Gamma_{\chi}^{(\phi)}(s;m_{12})| < 10^{-3}$ for such values of $g_{\phi}$. We also recognize that difference between $\sigma_{eff}(s)$ and $\sigma(s)$ at $s\ll\Lambda^2$ decreases for higher values of $\Lambda$ in the discussed limit of couplings magnitudes. This fact is explained by the Appelquist-Carazzone decoupling theorem.
	
	If $|g_{\phi}|$ is not small, then one-loop mixing term in $\sigma_{approx}(s)$ is significant with respect to the four-fermion interaction contribution there. Relation of these two terms absolute values is as follows:
	\[\frac{|\delta\mathcal{M}^{(mix)}(s)|}{|\delta\mathcal{M}^{(4-ferm)}(s)|} = \frac{2g_{\phi}^2|P(s)|}{\left|\left[s - \mu^2 - \Pi_{\phi\phi}(s)\right]\Gamma_{\chi}^{(eff)}(s;m_1)\Gamma_{\chi}^{(eff)}(s;m_2)\right|},\quad \Pi_{\phi\chi}(s) = g_{\phi}g_{\chi}P(s).\]
	Here $P(s)$ does not depend on any Yukawa couplings. It could be seen from this expression that the relation between $|\delta\mathcal{M}^{(mix)}(s)|$ and $|\delta\mathcal{M}^{(4-ferm)}(s)|$ does not depend on $g_{\chi}$ and is proportional to $g_{\phi}^2$. $|\delta\mathcal{M}^{(mix)}(s)|$ consists significant fraction of $|\delta\mathcal{M}^{(4-ferm)}(s)|$ near $s = \mu^2$ or if $|g_{\phi}|$ is big. For the model discussed here we have that $\left|\delta\mathcal{M}^{(mix)}(s)\right|\gtrsim 0.5\left|\delta\mathcal{M}^{(4-ferm)}(s)\right|$ for $\mu^2 < s \leq \Lambda^2$, $20\mu \leq \Lambda \leq 200\mu$ if $|g_{\phi}| \gtrsim 1$. Hence, if Yukawa coupling in the light sector of the model is strong enough, then one-loop mixing of light and heavy scalars should not be neglected in the process cross-section in the decoupling limit of the heavy boson.
	
	Since one-loop corrections in $\Gamma_{\chi}^{(eff)}(s;m_{12})$ are proportional to $g_{\phi}^2$, then they also should be taken into account if $|g_{\phi}|$ is big. In our model we have that $|\delta\Gamma_{\chi}^{(\phi)}(s;m_{12})|\gtrsim 0.1$ if $|g_{\phi}| > 1$.
	
	\section{The $t$-channel scattering process}
	\label{sec:t-channel-process-analyzis}
	
\begin{figure}[h]
		\centering
		\includegraphics[scale=0.4]{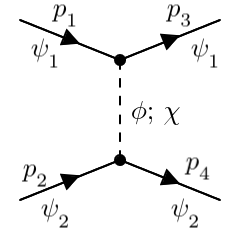}
		\caption{Diagram of the reaction $\psi_1 \psi_2 \rightarrow \psi_1 \psi_2$ within the UV-complete model described by the Lagrangian \eqref{full-model-lagrangian}}
		\label{fig:t-channel-process-uv-complete-diagram}
	\end{figure}

	We now consider a process $\psi_1\psi_2\rightarrow \psi_1\psi_2$, which takes place in $t$-channel only. Feynman diagram of this process is displayed in Fig.\,\ref{fig:t-channel-process-uv-complete-diagram}.
	Cross-section of this reaction within the UV-complete theory in the improved Born approximation is as follows:
	\begin{align}
	\label{t-channel-process-uv-complete-cross-section}
	\frac{d\sigma(\psi_1\psi_2 \rightarrow \psi_1\psi_2)}{d\Omega} &= \Phi^{(t)}(t)\left|\begin{pmatrix}
	-i g_{\phi}\Gamma_{\phi}(t; m_2) \\
	-i g_{\chi}\Gamma_{\chi}(t; m_2)
	\end{pmatrix}^T\right.\times \nonumber \\ 
	&\times\left.\begin{pmatrix}
	t - \mu^2 - \Pi_{\phi\phi}(t) & -\Pi_{\phi\chi}(t) \\
	-\Pi_{\phi\chi}(t) & t - \Lambda^2 - \Pi_{\chi\chi}(t)
	\end{pmatrix}^{-1} \begin{pmatrix}
	-i g_{\phi}\Gamma_{\phi}(t; m_1) \\
	-i g_{\chi}\Gamma_{\chi}(t; m_1)
	\end{pmatrix}\right|^2, \\
	\Phi^{(t)}(t) &= \frac{(4m_1^2 - t)(4m_2^2 - t)}{64\pi^2(E_1 + E_2)^2},\; t = (p_1 - p_3)^2 = 2\left(m_1^2 - E_1 E_3 + |\overline{p}_1||\overline{p}_3|\cos{\theta}\right), \nonumber \\
	d\Omega &= 2\pi\sin{\theta}d\theta,\; E_1 = \sqrt{\overline{p}_1^2 + m_1^2},\; E_2 = \sqrt{\overline{p}_2^2 + m_2^2},\; E_3 = \sqrt{\overline{p}_3^2 + m_1^2}. \nonumber
	\end{align}
	We omit box diagrams from our treatment of the $t$-channel process, too. It is assumed that their contributions are negligible, as it is for the $s$-process considered in section\,\ref{sec:s-channel-process-analyzis}.
	
	Expression \eqref{t-channel-process-uv-complete-cross-section} is also derived in the initial particles center-of-mass reference frame. Functions $\Gamma_{\phi}(t; m_{12})$ and $\Gamma_{\chi}(t; m_{12})$ represent Yukawa vertexes of $\phi$ and $\chi$ bosons with one-loop corrections. Diagrams of the latter are displayed in Fig.\,\ref{fig:yukawa-vertex-corrections-uv-complete-model}. Function $\Phi^{(t)}(t)$ in \eqref{t-channel-process-uv-complete-cross-section} is a kinematical factor. It is a product of two contributions. The first is introduced by integrations over the momentums of final particles. The second corresponds to  averaging and summation over spins orientations of initial and final particles.
	
\begin{figure}
		\centering
		\begin{subfigure}{0.5\textwidth}
			\centering
			\includegraphics[scale=0.35]{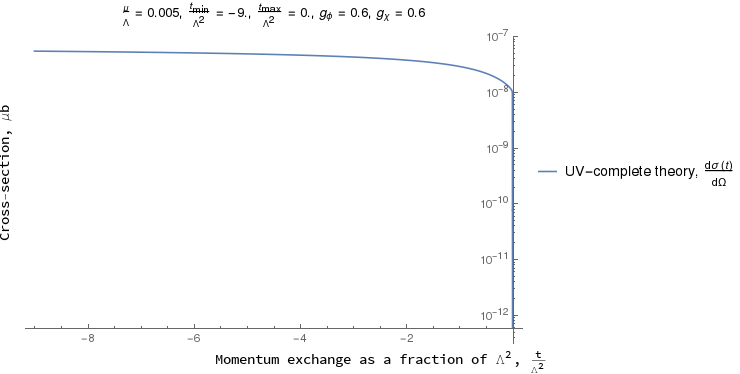}
			\caption{UV-complete theory}
			\label{fig:t-channel-process-uv-complete-theory-cross-section}
		\end{subfigure} \\
		\begin{subfigure}{0.5\textwidth}
			\centering
			\includegraphics[scale=0.35]{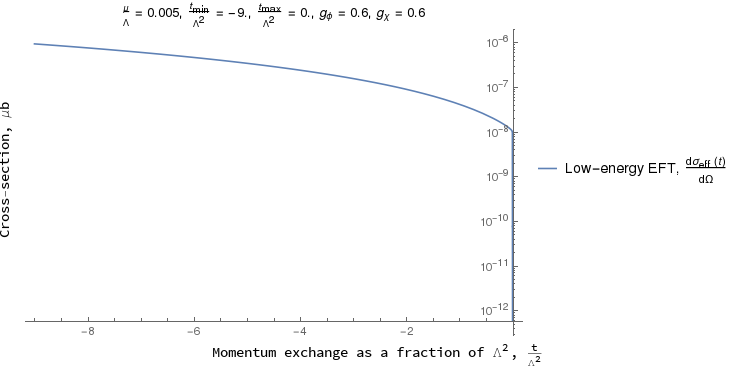}
			\caption{Low-energy EFT}
			\label{fig:t-channel-process-effective-theory-cross-section}
		\end{subfigure}
		\caption{Example cross-sections of the process $\psi_1\psi_2\rightarrow \psi_1\psi_2$}
		\label{fig:t-channel-process-typical-cross-sections}
	\end{figure}

	As it could be seen from the expression \eqref{t-channel-process-uv-complete-cross-section}, matrix element of the $t$-channel process has similar analytical structure as that of the $s$-channel process. That is, squared modulus factors in \eqref{s-channel-process-uv-complete-cross-section} and \eqref{t-channel-process-uv-complete-cross-section} are analytically similar and depend on Mandelstam invariants $s$ and $t$, accordingly. The same holds for the cross-section of this reaction $d\sigma_{eff}(\psi_1\psi_2 \rightarrow \psi_1\psi_2) / d\Omega$ within the low-energy EFT \eqref{effective-lagrangian} and approximate expression of the cross-section $d\sigma_{approx}(\psi_1\psi_2 \rightarrow \psi_1\psi_2) / d\Omega$ derived from \eqref{t-channel-process-uv-complete-cross-section} in the limit $|t| \ll \Lambda^2$. Hence, we do not provide expressions for $d\sigma_{eff}(\psi_1\psi_2 \rightarrow \psi_1\psi_2) / d\Omega$ and $d\sigma_{approx}(\psi_1\psi_2 \rightarrow \psi_1\psi_2) / d\Omega$ in this section. Examples of the $t$-channel process cross-sections are shown in Fig.\,\ref{fig:t-channel-process-typical-cross-sections}.
	
\begin{figure}
		\centering
		\includegraphics[scale=0.3]{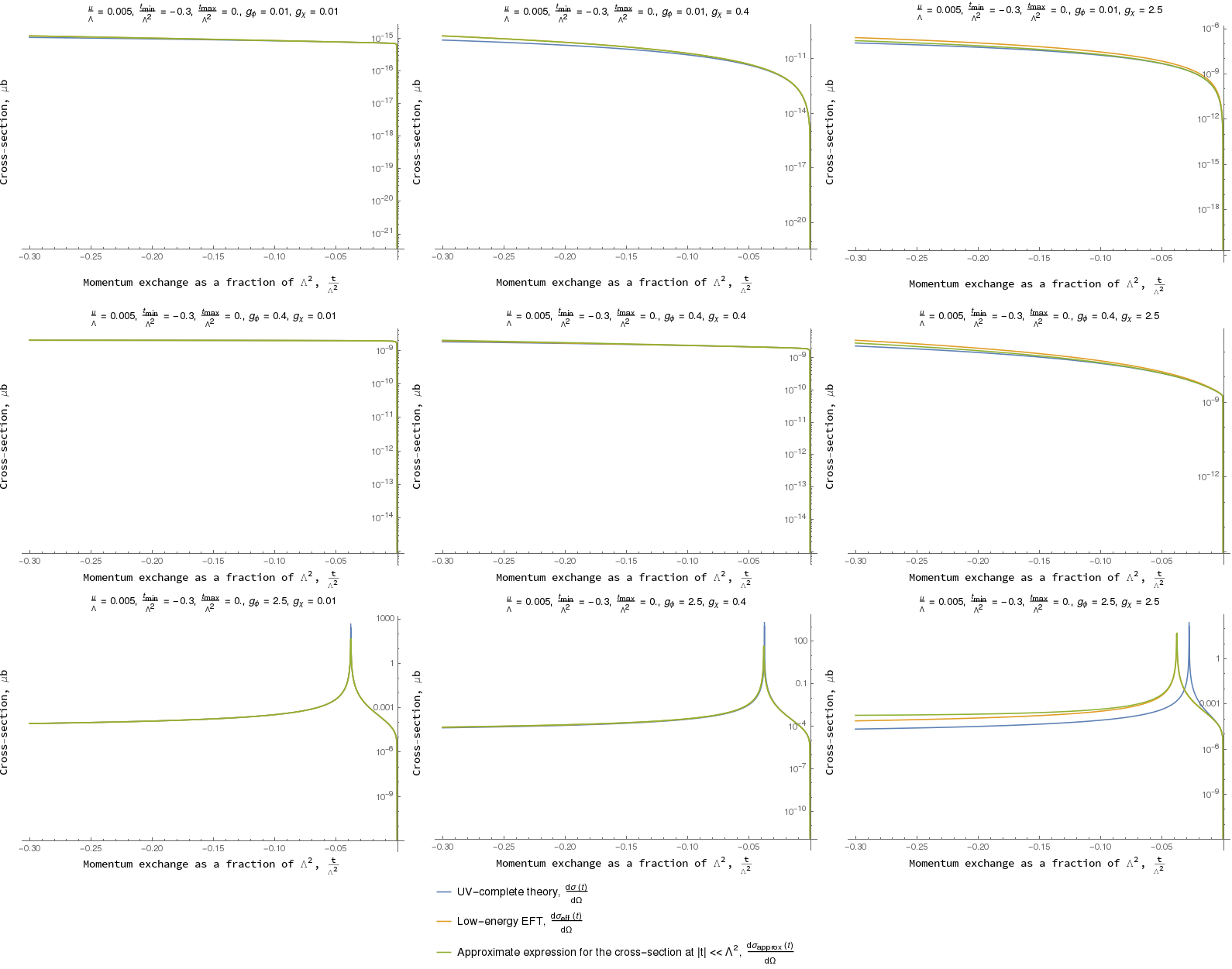}
		\caption{$\frac{d\sigma(t)}{d\Omega}$, $\frac{d\sigma_{eff}(t)}{d\Omega}$ and $\frac{d\sigma_{approx}(t)}{d\Omega}$ at $\Lambda = 200\mu$, $t$-channel process}
		\label{fig:t-process-cross-sections}
	\end{figure}

\begin{figure}
		\centering
		\includegraphics[scale=0.3]{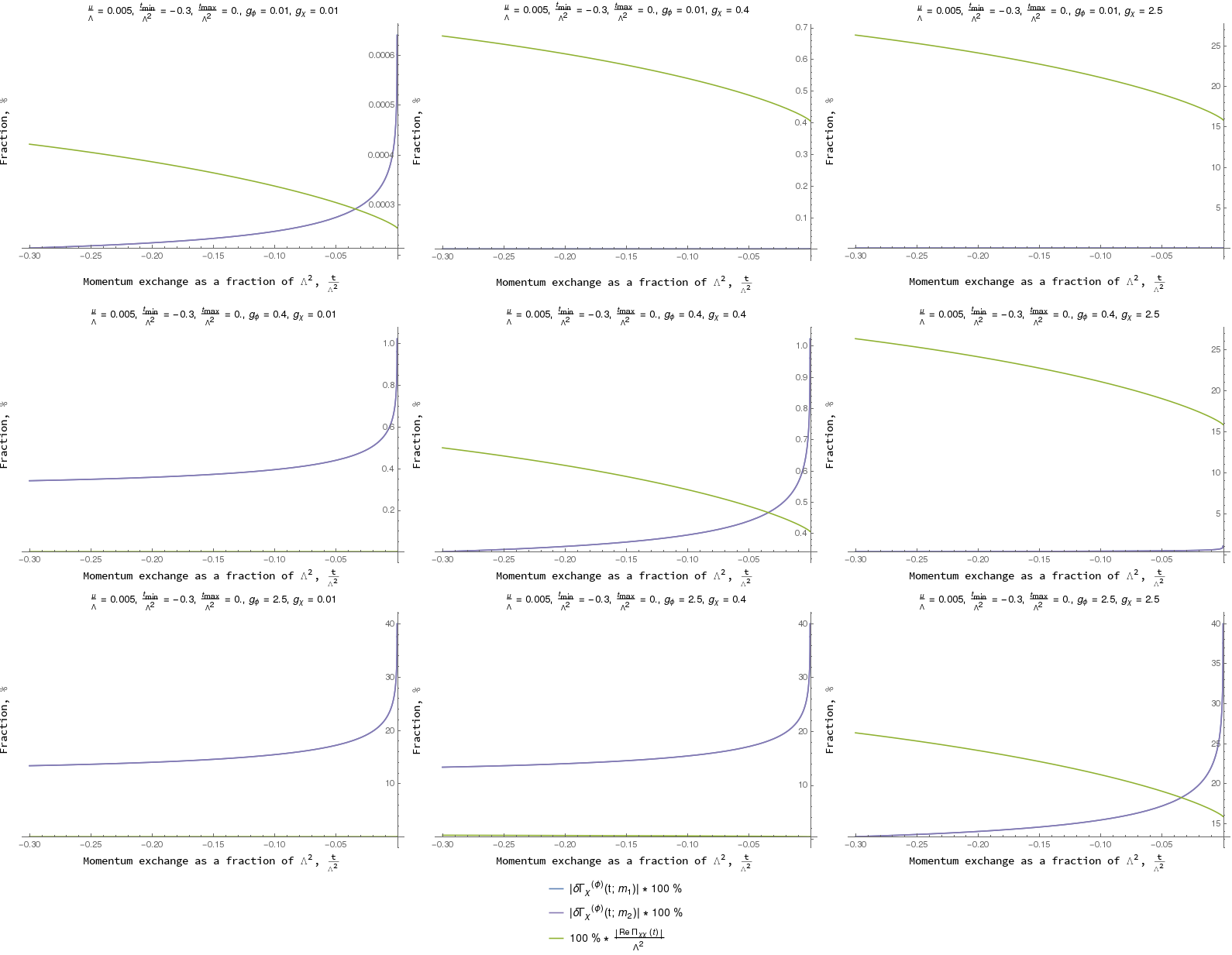}
		\caption{Magnitudes of loop corrections to Yukawa vertexes and mass of $\chi$ boson, $t$-channel process}
		\label{fig:t-process-loop-corrections-values}
	\end{figure}

\begin{figure}
		\centering
		\includegraphics[scale=0.3]{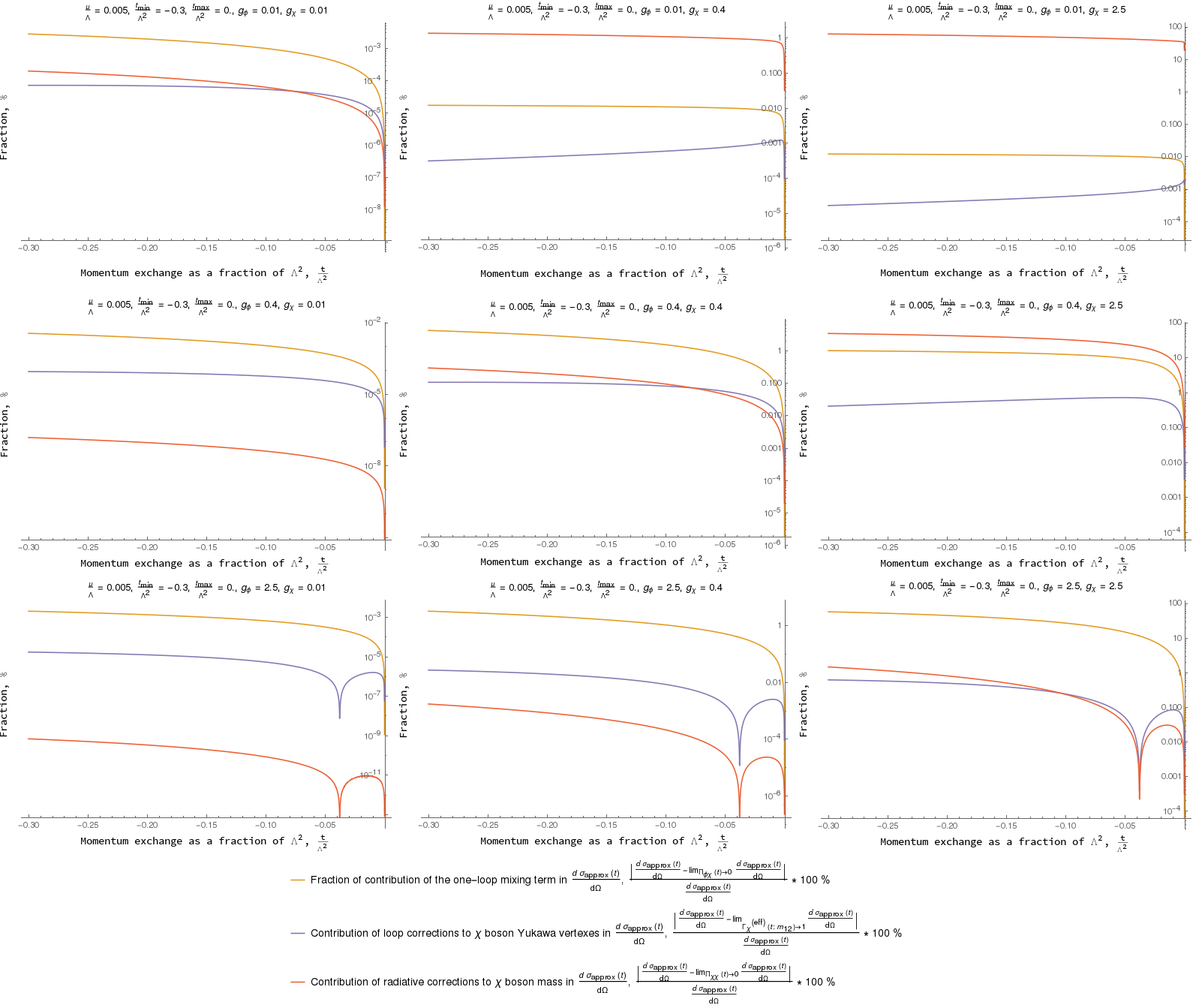}
		\caption{Relative contributions of various loop corrections in $\frac{d\sigma_{approx}(t)}{dt}$, $t$-channel process}
		\label{fig:t-process-loop-corrections-contributions}
	\end{figure}

	We carry out similar analysis for the scattering process $\psi_1\psi_2\rightarrow \psi_1\psi_2$ as it was established  out for the process $\psi_1\bar{\psi}_1\rightarrow \psi_2\bar{\psi}_2$ in section\,\ref{sec:s-channel-process-analyzis}. Corresponding plots are shown in Fig.\,\ref{fig:t-process-cross-sections}, \ref{fig:t-process-loop-corrections-values} and \ref{fig:t-process-loop-corrections-contributions}.
	
	Similarly to section\,\ref{sec:s-channel-process-analyzis}, we first identify a scenario when radiative corrections to couplings and mass of $\chi$ boson, as well as one-loop mixing contribution, are negligible at $|t|\ll \Lambda^2$. It could be seen in the graphs in Fig.\,\ref{fig:t-process-loop-corrections-values} that it is so if both $|g_{\phi}|$ and $|g_{\chi}|$ are small. In our model we have that if $|g_{\phi}| \lesssim 0.4$ and $|g_{\chi}|\lesssim 0.4$, then $|\Re\,\Pi_{\chi\chi}(t)| < 0.05 \Lambda^2$ and $|\delta\Gamma_{\chi}^{(\phi)}(t;m_{12})| < 0.05$ for $-1.5\Lambda^2 \leq t < 0$ and $20\mu \leq \Lambda \leq 200\mu$. $\delta\mathcal{M}^{(mix)}$ is also small, such that $|\delta\mathcal{M}^{(mix)}| < 0.15 |\delta\mathcal{M}^{(4-ferm)}|$ in the same ranges of $t$ and $\Lambda$. All these corrections contribute less than $6\,\%$ of $d\sigma_{approx} / d\Omega$ for $-0.3\Lambda^2 \leq t < 0$ and $20\mu \leq \Lambda \leq 200\mu$. This could be observed in Fig.\,\ref{fig:t-process-loop-corrections-contributions} for $\Lambda = 200\mu$.
	
	Now we proceed to scenarios when radiative corrections are considerable.
	
	If $|g_{\chi}|$ is big, then $|\Pi_{\chi\chi}(t)|$ is significant with respect to $\Lambda^2$. In our model we have that $|\Pi_{\chi\chi}(t)| \gtrsim 0.1\Lambda^2$ at $|t| \ll \Lambda^2$ if $2 \leq |g_{\chi}| \leq 3$. According to plots in Fig.\,\ref{fig:t-process-loop-corrections-contributions}, this correction contributes more than $10\,\%$ of $d\sigma_{approx}(t) / d\Omega$ for $-0.3\Lambda^2\leq t < 0$ if $0.01\leq |g_{\phi}|\leq 0.05$ -- in this limit all other corrections are negligible. This takes place for $20\mu\leq \Lambda \leq 200\mu$. Thus, for such choice of the model parameters loop correction to heavy boson mass becomes significant even at low $|t|$.
	
	If $|g_{\phi}|$ is big, then loop corrections to $\chi$ boson Yukawa vertexes are considerable. In present model we have that if $|g_{\phi}|\gtrsim 1.5$ then $|\delta\Gamma_{\chi}^{(\phi)}(t;m_{12})| \geq 0.1$ at $|t|\ll \Lambda^2$. This is so for $20\mu\leq \Lambda \leq 200\mu$. The one-loop mixing term is also significant in the discussed limit. Namely, we have that $|\delta\mathcal{M}^{(mix)}| \geq 0.1|\delta\mathcal{M}^{(4-ferm)}|$ if $|g_{\phi}| \geq 0.8$ and $|t|\ll \Lambda^2$. If $|g_{\phi}|$ is very big, then $|\delta\mathcal{M}^{(mix)}| > |\delta\mathcal{M}^{(4-ferm)}|$ even at small $|t|$. We find out that it is so if $|g_{\phi}| \geq 1.5$ and $100\mu \leq \Lambda \leq 200\mu$. If $\Lambda = 20\mu$, then $|\delta\mathcal{M}^{(mix)}|$ is bigger than $|\delta\mathcal{M}^{(4-ferm)}|$ at small $|t|$ if $|g_{\phi}|\geq 2$. Thus, if Yukawa interaction in the light sector of the model is strong enough, then one-loop mixing of light and heavy scalars is significant even at low $|t|$.
	
	It could be observed in the cross-sections plots in Fig.\,\ref{fig:t-process-cross-sections} that if $|g_{\phi}|$ is big, then $d\sigma(t) / d\Omega$ develops a peak at some $t = t_0$. This peak corresponds to the point where determinant of the matrix in \eqref{t-channel-process-uv-complete-cross-section} is zero and $d\sigma(t) / d\Omega$ apparently diverges. That is, $t_0$ satisfies the following equation:
	\[\left[t_0 - \mu^2 - \Pi_{\phi\phi}(t_0)\right]\left[t_0 - \Lambda^2 - \Pi_{\chi\chi}(t_0)\right] - \Pi_{\phi\chi}^2(t_0) = 0.\]
	This equation contains terms of the fourth order in Yukawa couplings. Two-loop radiative corrections enter perturbative expansions at this order, too. Hence, the behaviour of $d\sigma(t) / d\Omega$ near the point $t = t_0$ could be studied only when two-loop diagrams are taken into account in the expressions for $\Pi_{\phi\phi}(t)$, $\Pi_{\phi\chi}(t)$ and $\Pi_{\chi\chi}(t)$. Such analyzis is beyond the scope of this paper, so we omit it for now.
	
	\section{Discussion and conclusion}
	\label{sec:conclusion}
	
	In this paper we derived the low-energy effective Lagrangian of generalized Yukawa model in the limit when the heaviest scalar field of the model decouples. There are two scalars in the model -- $\phi$ and $\chi$, which are light and heavy, respectively.
	
	We analyzed contributions of corrections from loops with light particles in the cross-sections of scattering processes within the model in the limit when $\chi$ decouples. Two reactions were considered -- $\psi_1\bar{\psi}_1 \rightarrow \psi_2\bar{\psi}_2$ and $\psi_1\psi_2\rightarrow \psi_1\psi_2$. These processes take place in $s$- and $t$-channel, respectively. We identified values of the model Yukawa couplings when radiative corrections are insignificant at low energies and effective Lagrangian \eqref{effective-lagrangian} is valid.
	
	We found out that if $|g_{\phi}|$ and $|g_{\chi}|$ are small the loop corrections are negligible and the EL \eqref{effective-lagrangian} is applicable for description of the scattering processes at low energies. For the model parameters values in table\,\ref{tab:fixed-parameters-values} it is so if $|g_{\phi}| < 0.4$ and $|g_{\chi}| < 0.4$.
	
	If $|g_{\chi}|$ is big, then loop corrections to $\chi$ boson mass are to be considerable even at low energies. In our model, we have that if $|g_{\chi}| > 2.5$ then corrections displayed in Fig.\,\ref{fig:chi-chi-polarization-operator} consist more than $10\,\%$ of the $\chi$ boson mass. These corrections suppress the contribution of the heavy scalar in a reaction cross-section. Hence, if $|g_{\chi}|$ is big, then \eqref{effective-lagrangian} significantly overestimates $\chi$ boson contribution in a cross-section when $\chi$ decouples.
	
	If $|g_{\phi}|$ is not small, then contribution of the scalar fields one-loop mixing is significant in both $s$- and $t$-channels. One-loop mixing of $\phi$ and $\chi$ is introduced by the diagram in Fig.\,\ref{fig:phi-chi-polarization-operator}. In our model, we get that if $|g_{\phi}| \gtrsim 1.5$, then modulus of the contribution of scalars one-loop mixing in the matrix elements of the considered reactions is bigger than $50\,\%$ of the four-fermion interaction term modulus at $s \gtrsim \mu^2$ and $|t| \gtrsim \mu^2$. This was observed in the limit when $|\Pi_{\chi\chi}(p^2)|$ is negligible. Radiative corrections to the Yukawa vertexes of $\chi$ boson are also significant in this limit, since they are proportional to $g_{\phi}^2$. These corrections are shown in Fig.\,\ref{fig:phi-loop-in-yukawa-vertex}.
	
	To conclude, in our investigation we have derived the conditions when the radiative corrections are negligible in the decoupling limit for the cross-sections of reactions within the generalized Yukawa model. According to these conditions, radiative corrections become significant if the interactions between light fermions and either light or decoupled scalar are not feeble. In such scenarios, expression \eqref{effective-lagrangian} should not be used as the low-energy approximation of the model \eqref{full-model-lagrangian}. The results obtained in this investigation could be applied to some models of new physics which extend the SM, such as the two-Higgs-doublet model (2HDM). The latter has a limit when new heavy particles decouple. So, low-energy EL could be derived for it \cite{bibl:dmytriiev-skalozub-eff-lagrangian}. Contributions of radiative corrections into cross-sections within the 2HDM should be estimated in the limit when heavy fields beyond the SM decouple. This is the problem left for the future.

\end{document}